\documentclass{webofc}
\usepackage[varg]{txfonts}   
\begin{document}
\title{Cluster cosmology with the NIKA2 SZ Large Program}
%
%

\author{\firstname{F.} \lastname{Mayet} \inst{\ref{LPSC}}\fnsep\thanks{\email{mayet@lpsc.in2p3.fr}}
\and \firstname{R.} \lastname{Adam} \inst{\ref{LLR},\ref{CEFCA}}
\and  \firstname{P.} \lastname{Ade} \inst{\ref{Cardiff}}
\and  \firstname{P.} \lastname{Andr\'e} \inst{\ref{CEA1}}
\and  \firstname{A.} \lastname{Andrianasolo} \inst{\ref{IPAG}}
\and \firstname{M.} \lastname{Arnaud} \inst{\ref{CEA1}}
\and  \firstname{H.} \lastname{Aussel} \inst{\ref{CEA1}}
\and \firstname{I.} \lastname{Bartalucci} \inst{\ref{CEA1}}
\and  \firstname{A.} \lastname{Beelen} \inst{\ref{IAS}}
\and  \firstname{A.} \lastname{Beno\^it} \inst{\ref{Neel}}
\and  \firstname{A.} \lastname{Bideaud} \inst{\ref{Neel}}
\and  \firstname{O.} \lastname{Bourrion} \inst{\ref{LPSC}}
\and  \firstname{M.} \lastname{Calvo} \inst{\ref{Neel}}
\and  \firstname{A.} \lastname{Catalano} \inst{\ref{LPSC}}
\and  \firstname{B.} \lastname{Comis} \inst{\ref{LPSC}}
\and  \firstname{M.} \lastname{De~Petris} \inst{\ref{Roma}}
\and  \firstname{F.-X.} \lastname{D\'esert} \inst{\ref{IPAG}}
\and  \firstname{S.} \lastname{Doyle} \inst{\ref{Cardiff}}
\and  \firstname{E.~F.~C.} \lastname{Driessen} \inst{\ref{IRAMF}}
\and  \firstname{A.} \lastname{Gomez} \inst{\ref{CAB}}
\and  \firstname{J.} \lastname{Goupy} \inst{\ref{Neel}}
\and  \firstname{F.} \lastname{K\'eruzor\'e} \inst{\ref{LPSC}}
\and  \firstname{C.} \lastname{Kramer} \inst{\ref{IRAME}}
\and  \firstname{B.} \lastname{Ladjelate} \inst{\ref{IRAME}}
\and  \firstname{G.} \lastname{Lagache} \inst{\ref{LAM}}
\and  \firstname{S.} \lastname{Leclercq} \inst{\ref{IRAMF}}
\and  \firstname{J.-F.} \lastname{Lestrade} \inst{\ref{LERMA}}
\and  \firstname{J.F.} \lastname{Mac\'ias-P\'erez} \inst{\ref{LPSC}}
\and  \firstname{P.} \lastname{Mauskopf} \inst{\ref{Cardiff},\ref{Arizona}}
\and  \firstname{A.} \lastname{Monfardini} \inst{\ref{Neel}}
\and  \firstname{L.} \lastname{Perotto} \inst{\ref{LPSC}}
\and  \firstname{G.} \lastname{Pisano} \inst{\ref{Cardiff}}
\and \firstname{E.} \lastname{Pointecouteau} \inst{\ref{IRAP}, \ref{IRAP2}}
\and  \firstname{N.} \lastname{Ponthieu} \inst{\ref{IPAG}}
\and \firstname{G.~W.} \lastname{Pratt} \inst{\ref{CEA1}}
\and  \firstname{V.} \lastname{Rev\'eret} \inst{\ref{CEA1}}
\and  \firstname{A.} \lastname{Ritacco} \inst{\ref{IRAME}}
\and  \firstname{C.} \lastname{Romero} \inst{\ref{IRAMF}}
\and  \firstname{H.} \lastname{Roussel} \inst{\ref{IAP}}
\and  \firstname{F.} \lastname{Ruppin} \inst{\ref{MIT}}
\and  \firstname{K.} \lastname{Schuster} \inst{\ref{IRAMF}}
\and  \firstname{S.} \lastname{Shu} \inst{\ref{IRAMF}}
\and  \firstname{A.} \lastname{Sievers} \inst{\ref{IRAME}}
\and  \firstname{C.} \lastname{Tucker} \inst{\ref{Cardiff}}
\and  \firstname{R.} \lastname{Zylka} \inst{\ref{IRAMF}}
}

\institute{\label{LPSC} Univ. Grenoble Alpes, CNRS, Grenoble INP, LPSC-IN2P3, 53, av. des Martyrs, 38000 Grenoble, France
\and \label{LLR} LLR, CNRS, \'Ecole Polytechnique, Institut Polytechnique de Paris, Palaiseau, France  
\and \label{CEFCA} Centro de Estudios de Fisica del Cosmos de Aragon (CEFCA), Plaza San Juan, 1, planta 2, E-44001, Teruel, Spain 
\and \label{Cardiff} Astronomy Instrumentation Group, University of Cardiff, UK          
\and \label{CEA1} AIM, CEA, CNRS, Universit\'e Paris-Saclay, Universit\'e Paris Diderot, Sorbonne Paris Cit\'e, 91191 Gif-sur-Yvette, France    
\and \label{IPAG} Univ. Grenoble Alpes, CNRS, IPAG, 38000 Grenoble, France     
\and \label{IAS} Institut d'Astrophysique Spatiale (IAS), CNRS and Universit\'e Paris Sud, Orsay, France    
\and \label{Neel} Institut N\'eel, CNRS and Universit\'e Grenoble Alpes, France
\and \label{Roma} Dipartimento di Fisica, Sapienza Universit\`a di Roma, Piazzale Aldo Moro 5, I-00185 Roma, Italy       
\and \label{IRAMF} Institut de RadioAstronomie Millim\'etrique (IRAM), Grenoble, France 
\and \label{CAB} Centro de Astrobiolog\'ia (CSIC-INTA), Torrej\'on de Ardoz, 28850 Madrid, Spain
\and \label{IRAME} Instituto de Radioastronom\'ia Milim\'etrica (IRAM), Granada, Spain 
\and \label{LAM} Aix Marseille Univ, CNRS, CNES, LAM, Marseille, France
\and \label{LERMA} LERMA, Observatoire de Paris, PSL Research University, CNRS, Sorbonne Universit\'es, UPMC Univ. Paris 06, 75014 Paris, France
\and \label{Arizona} School of Earth and Space Explo. and Dep. of Physics, Arizona State University, Tempe, AZ 85287         
\and \label{IRAP} Univ. de Toulouse, UPS-OMP, Institut de Recherche en Astrophysique et Plan\'etologie, Toulouse, France
\and \label{IRAP2} CNRS, IRAP, 9 Av. colonel Roche, BP 44346, F-31028 Toulouse cedex 4, France  
\and \label{IAP} Institut d'Astrophysique de Paris, CNRS (UMR7095), 98 bis boulevard Arago, 75014 Paris, France
\and \label{MIT} Kavli Institute for Astrophysics and Space Research, Massachusetts Institute of Technology, Cambridge, MA 02139, USA 
          }
	  
\abstract{The main limiting factor of cosmological analyses based on thermal Sunyaev-Zel'dovich (SZ) cluster statistics comes from the bias and systematic uncertainties that affect the estimates of the 
mass of galaxy clusters. High-angular resolution SZ observations at high redshift are needed to study a potential redshift or morphology dependence of both the mean pressure profile and of the 
mass-observable scaling relation used in SZ cosmological analyses.
The NIKA2 camera is a new generation continuum instrument installed at the IRAM 30-m telescope. 
With a large field of view, a high angular resolution and a high-sensitivity, the NIKA2 camera has unique SZ mapping capabilities. In this paper, we present the NIKA2 SZ large program, 
aiming at observing a large sample of clusters at redshifts between $0.5$ and $0.9$, and the characterization of the first cluster oberved with NIKA2.}
%
\maketitle
\section{Cluster cosmology: the need for high-angular resolution SZ observation}
\label{sec:cosmo}
There is an increasing body of evidence of discrepancies between cosmological estimations obtained with early and late Universe probes \cite{Verde:2019ivm,Wong:2019kwg}. In particular, a tension 
is observed  between CMB and cluster-derived cosmological parameters, see e.g. \cite{Ade:2015fva}. If confirmed, this may be a sign of new physics, related with {\it e.g.} structure formation scenario and neutrino physics \cite{Bolliet:2019zuz}. It may also be
explained by an insufficient knowledge of cluster physics, and in particular of the main tools that are used to extract cosmological constraints from a  sample of clusters of galaxies, namely: the hydrostatic bias parameter \cite{Salvati:2019zdp}, the mean pressure profile \cite{Ruppin:2019deo,Ruppin:2019mcb} and the mass-observable scaling relation \cite{Pratt:2019cnf}.\\
Clusters are inherently multi-wavelength objects. Amongst various probes, the thermal Sunyaev-Zel'dovich (SZ)  
effect plays a key role. It is the inverse Compton scattering of CMB photons on hot electrons of the intra-cluster medium (ICM) \cite{Sunyaev:1972}. It induces a shift of the CMB black-body 
spectrum to higher frequency, with a decrease of the CMB intensity below $217 \ {\rm GHz}$ and an increase above. The SZ effect is thus a spectral distortion of the 
CMB spectrum. This is the reason why it is redshift-independent and can thus be used for the observation of high-redshift clusters \cite{Mroczkowski:2018nrv}. 
Clusters of galaxies are known to be powerful probes to study cosmology as their number and distribution in mass and redshift is dependent on the geometry of the Universe. Cosmological parameters can be extracted either from a number count on a cluster catalog, per redshift and mass bin,  or from SZ power spectrum 
analyses performed on SZ sky map \cite{Ade:2015fva,Aghanim:2015eva}.\\ 
The mass-observable scaling relation is needed to relate the cluster mass to the SZ signal  given by the integrated Compton parameter $Y_{500}$, which is the spherical integration of the electronic pressure 
up to a radius\footnote{$R_{500}$ is defined as the radius for which the mean cluster density is $500$ times the critical density.} $R_{500}$. 
For a cosmological survey, $Y_{500}$ is the only information available and  it must be related to the cluster mass via the scaling relation. In particular, the {\it Planck} collaboration published a 
scaling relation calibrated with $71$ clusters, at redshift below $0.45$, for which the masses have been obtained from X-ray observations only \cite{Planck-scaling}. This scaling relation may be 
applied to the Planck-cluster population but it may depend on the redshift range, the cluster dynamical state ({\it e.g.} mergers, overpressures, ...) and the cluster morphology  ({\it e.g.} departure from sphericity).\\
The mean pressure profile is needed for instance in cluster count analyses to measure the integrated SZ flux of each cluster \cite{Melin:2006qq,Planck-pressure}, when the angular resolution 
of current SZ surveys does not enable the measurement of pressure profiles for individual clusters. In the self-similarity hypothesis, clusters of galaxies are supposed to be a scaled version of each other. Hence, the mean pressure profile may be applied to the whole population. 
Currently, the most widely used mean profile has been evaluated  in \cite{Arnaud2010} with $33$ X-ray-selected clusters at low redshift ($z<0.2$) and at high mass ($>10^{14} \ M_{\odot}$), together with
simulated ones. All observed profiles have been obtained with X-ray information only (spectroscopy). 
As for the scaling relation, it may depend on redshift, dynamical state and morphology \cite{Arnaud2010}. Moreover, one could question the validity of the self-similarity 
hypothesis at high redshift where clusters are encountering merging events \cite{sembo}.\\
To assess results in cluster cosmology, 
these two tools must be studied with SZ observations  on large range of cluster masses ($10^{14}-10^{15} \ M_{\odot}$) and at high redshift $z>0.5$ for which X-ray observations become time expensive. 
Detailed information on the intra-cluster medium is indeed needed to study the impact of dynamical state and morphology. Both the calibration precision and scatter of the mass-observable scaling 
relation and of the mean pressure profile must be carefully studied in order to be included in forthcoming cluster cosmology analyses.\\
One may write down an instrument wish list for SZ science aiming at cluster cosmology. It includes: 
high angular resolution, in order to resolve inner structures, high sensitivity, to reduce integration time, large field of view, to map the cluster up to its outskirts, and dual band observation, to remove
point-source contamination. 
These requirements correspond to the performance and characterics of the NIKA2 camera at the IRAM 30-m telescope \cite{Calvo2016,NIKA2-Adam,Perotto2019} that makes it a powerful tool for SZ observations.
 
\section{The NIKA2 SZ large program}
\label{sec:lpsz}

\begin{figure}[t]
\begin{center}
\includegraphics[scale=0.65]{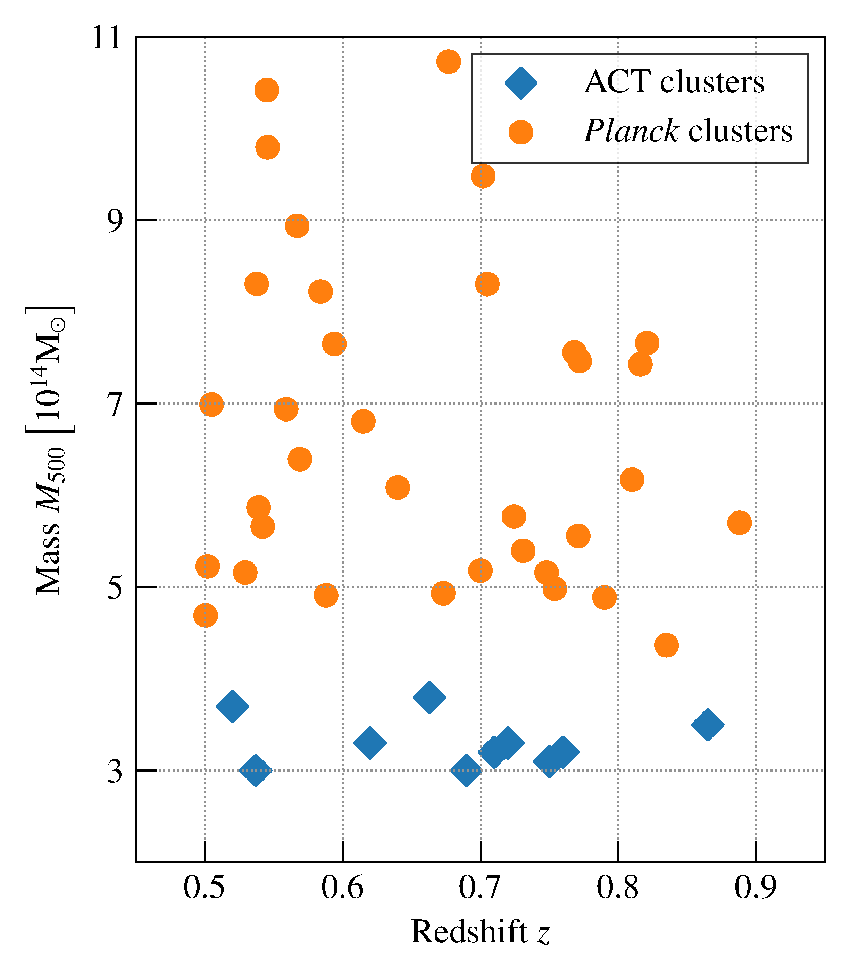}
\caption{The NIKA2 LPSZ cluster sample presented in the mass-redshift plane, see text for details.}
\label{fig:sampleLPSZ}     
\end{center}
\end{figure}
The NIKA2 SZ large program (hereafter LPSZ) is part of the NIKA2 guaranteed time, allocated by the IRAM scientific committee to the NIKA2 consortium. $300$ hours of observations have been granted for 50 clusters that will be observed with NIKA2 at the IRAM 30-m telescope in the 
forthcoming years (up to 2022).\\ 
The immediate objective of the NIKA2 SZ large program is to study the thermodynamical properties of a representative sample of SZ-selected (from {\it Planck} and ACT samples) clusters of galaxies at intermediate and high redshift ($0.5< z < 0.9$) spanning one order of magnitude in mass. The high-resolution follow-up of SZ-discovered clusters will allow us, in combination 
with XMM-{\it Newton} X-ray data, to provide high quality deliverables in terms of radial profiles (density, pressure, mass, temperature, entropy). NIKA2 and {\it Planck} data will be used to 
extract the pressure profile $P(r)$ whereas XMM-{\it Newton} data enable the evaluation of the density profile $n(r)$. They are then combined to obtained the hydrostatic mass profile $M(r)$ but also the temperature profile ${\rm k_B}T(r) $ and the {\em entropy} profile
$K(r)$. 
All thermodynamic profiles are essential for a full understanding of the mean pressure profile and of the scaling relation in order to probe cluster physics and leverage 
large survey samples to constrain cosmology.\\ 
To define the sample, we followed the approach used in X-rays with the local reference REXCESS sample at $z<0.2$  \cite{Arnaud2010}, dedicated to the in-depth study of a representative sample of 33 X-ray selected clusters.
The NIKA2 SZ large program uniquely exploits the 
excellent match in sensitivity and spatial resolution 
of XMM-{\it Newton} and the NIKA2 camera. In particular, 
spatially-resolved X-ray and SZ pressure profiles can be 
compared to assess effects such as 3D structure or clumping. The full sample will allow us to undertake an unprecedented study at $z>0.5$ of the mean pressure profile and of the 
scaling relations, together with their dispersion. The sample will serve as an important probe of the physics of gravitational collapse and of the influence of non-gravitational processes on the ICM thermodynamics.\\
The cluster sample of the NIKA2 SZ large program currently contains $45$ SZ-selected clusters, amongst  which $10$ are from the ACT catalog and $35$ from the 
{\it Planck} one, see \cite{Ade:2015gva, Ade:2013skr,Hasselfield:2013wf}. The cluster sample is presented on 
figure~\ref{fig:sampleLPSZ} in the mass-redshift plane, where masses are $M_{500}$ from the Planck and ACT catalogues.
ACT clusters 
are located in the low mass range (below $4 \times 10^{14}\ M_{\odot}$). We have defined two bins in redshift and $5$ in mass. Within each bin we have selected $5$ clusters maximising, when possible, the
overlap with the SZ clusters observed or already planned for follow-ups with the XMM-{\it Newton} satellite.\\
One has to ensure that the sample is representative of the cluster population, {\it i.e.} not biased towards a given cluster morphology population. 
This condition will allow us to derive mass-observable scaling relations that can be applicable to the whole cluster population (not only relaxed or unrelaxed ICM) and to 
achieve a global characterization of clusters and an improved control of systematics due to their astrophysics. A flux-selected subset of a SZ-selected cluster catalogue fulfills the requirement of a representative sample.  We have considered the following main target selection criteria:
\begin{itemize}
\item clusters belonging to SZ-selected samples for which the redshift information is available,
\item $0.5 < z < 0.9$, to explore the cluster statistical properties beyond the local Universe,
\item $dec >  -11^\circ$, to ensure observability of the sources from the IRAM 30-m telescope.
\end{itemize}
Observation times have been chosen in order to have an homogeneous data quality, the criteria being  $S/N=3$ on the pressure profile at $R_{500}$. Most clusters already have X-ray data and follow-ups are in progress. {\it In fine}, all clusters data, SZ and X-ray, will be combined to evaluate thermodynamic profiles, which are the main deliverables of the project.\\
In terms of observations, careful treatment must be paid to data analysis since the SZ emission from clusters is both very faint and extended with respect to the NIKA2 beam. Point source contamination is also an issue worth mentioning but the NIKA2 dual-band observation should enable for self consistent foreground source subtraction.\\
We have demonstrated, through pilot studies conducted with the 
pathfinder NIKA \citep{Catalano:2014nml}, the  possibility to 
recover cluster thermodymanic profiles from the combination of 
NIKA SZ and XMM-{\it Newton} X-ray observations
 \cite{Adam:2013ufa, Adam:2014wxa, Adam:2015bba,Adam:2016abn,
 Ruppin:2016rnt, Adam:2017mlj,Romero:2017xri,Adam:2017zpu}.

\section{First cluster observation with NIKA2}
\label{sec:first}

\begin{figure}[t]
\begin{center}
\includegraphics[scale=0.4]{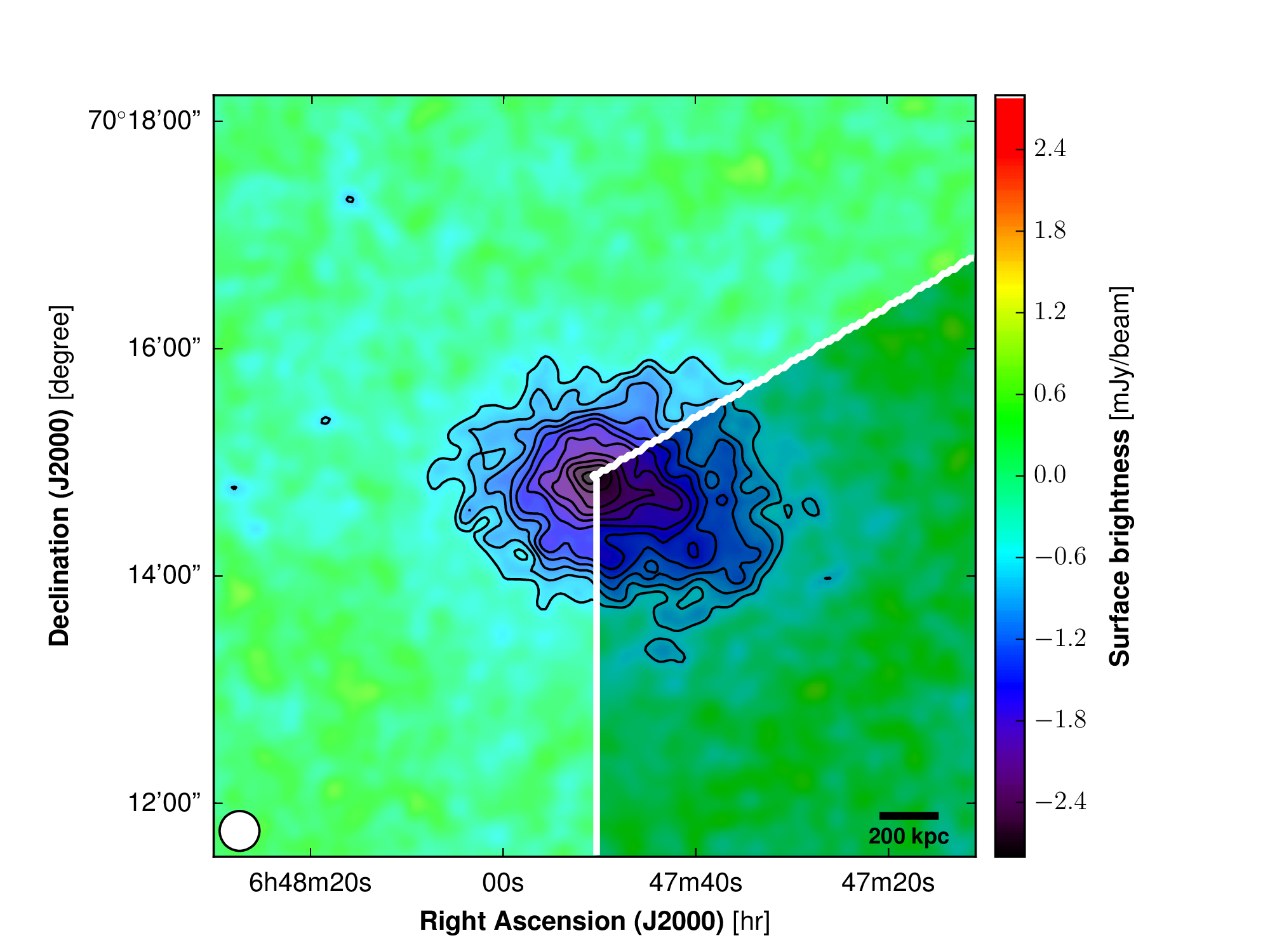}
\caption{NIKA2 tSZ surface brightness map at $150 \ {\rm GHz}$ after subtraction of the submillimeter point source contaminant. 
The over-pressure region is shown as a dark sector. Figure extracted from \cite{Ruppin:2017bnt}.}
\label{fig:144}       
\end{center}
\end{figure}

The first cluster of the LPSZ sample that has been observed  with NIKA2 is PSZ2 G144.83+25.11 at redshift $z = 0.58$ \cite{Ruppin:2017bnt}. 
This cluster has been chosen to begin the NIKA2 SZ Large program because it is expected to have a strong SZ signal. Moreover, NIKA2 data  may be combined with X-ray data as we have for this cluster deep 
XMM-{\it Newton} data ($60 \ {\rm ks}$). This allowed us to compare thermodynamic profiles obtained with SZ  and X-ray data to the ones obtained with X-ray only (with spectroscopy).
This cluster has been observed with an effective time of 11 hours with a mean atmospheric opacity of $0.3$ at $2 \ {\rm mm}$. At $2 \ {\rm mm}$, the SZ peak is detected at $13.5 \sigma$ and the SZ signal 
extends up to $1.4 \ {\rm arcmin}$, with a noise at the level of $\sim 200 \
{\rm \mu Jy/beam}$. At $1 \ {\rm mm}$, no SZ signal is detected, as expected
given the noise level ($\sim 930 \ {\rm \mu Jy/beam}$). Nonetheless, the  $1 \
{\rm mm}$ map is used to identify one point source that compensates the SZ
signal at $2 \ {\rm mm}$. The flux of this source at $2 \ {\rm mm}$ is estimated
from the fit of its Spectral Energy Distribution evaluated with {\it Herchel} data \cite{Herschel-1,Herschel-2} and NIKA2 (1 mm). The map is then corrected from this contamination. 
As shown on figure \ref{fig:144}, we have identified a thermal pressure excess in the south-west region of this cluster.\\
NIKA2 data have been used jointly with SZ data from  MUSTANG, Bolocam, and {\it Planck} experiments in order to non-parametrically set the best constraints on the electronic pressure
distribution from the cluster core ($R\sim 0.02R_{500}$) to its outskirts ($R\sim 3R_{500}$). We investigated the impact of the over-pressure region on the shape of the pressure profile and on the constraints on the integrated Compton parameter $Y_{500}$. A hydrostatic mass analysis has also been performed by combining the SZ-constrained pressure profile with the deprojected electronic density profile from XMM-{\it Newton}. This allowed us to conclude that the estimates of $Y_{500}$ and $M_{500}$ obtained from the analysis with and 
without masking the disturbed ICM region differ by $65\ \%$ and $79\ \%$ respectively.\\  
This work highlighted the fact that NIKA2 will have a crucial impact on the characterization
of the scatter of the $Y_{500}-M_{500}$ scaling relation due to its high potential to constrain the thermodynamic and morphological properties of the ICM when used in synergy with X-ray observations of similar angular resolution. 
This study also presents the typical products that will be delivered to the community for all clusters included in the NIKA2 SZ Large Program. Note that most clusters will not exhibit such a strong SNR, not
have Mustang/Bolocam data. 
The observation of the second  
cluster of the LPSZ is presented in \cite{Keruzore2019}.

\section*{Acknowledgements}
{\small
We would like to thank the IRAM staff for their support during the campaigns. The NIKA dilution cryostat has been designed and built at the Institut N\'eel. In particular, 
we acknowledge the crucial contribution of the Cryogenics Group, and in particular G. Garde, H. Rodenas, J. P. Leggeri, P. Camus. This work has been partially funded by 
the Foundation Nanoscience Grenoble and the LabEx FOCUS ANR-11-LABX-0013. This work is supported by the French National Research Agency under the contracts "MKIDS", "NIKA" and 
ANR-15-CE31-0017 and in the framework of the "Investissements d'avenir" program (ANR-15-IDEX-02). This work has benefited from the support of the European Research 
Council Advanced Grant ORISTARS under the European Union's Seventh Framework Programme (Grant Agreement no. 291294). 
We acknowledge fundings from the ENIGMASS French LabEx (R. A. and F. R.), the CNES post-doctoral fellowship program (R. A.), the CNES doctoral fellowship program (A. R.) 
and the FOCUS French LabEx doctoral fellowship program (A. R.). R.A. acknowledges support from Spanish Ministerio de Econom\'ia and Competitividad (MINECO) through grant number AYA2015-66211-C2-2.

}


\end{document}